\documentclass[12pt,onecolumn]{article}
\usepackage{amssymb,graphicx,url}
\usepackage{graphicx}

\author{Frank Nielsen,\\
Sony Computer Science Laboratories, Inc.\\
3-14-13 Higashi Gotanda, 141-0022 Shinagawa-ku, Tokyo, Japan\\
{\tt Frank.Nielsen@acm.org}
}

\title{Logging safely in public spaces using color PINs}

\begin{document}

\maketitle

\begin{abstract}
Nowadays, we are increasingly logging on many different Internet sites to access private data like emails or photos remotely stored in the clouds.
This makes us all the more concerned with digital identity theft and passwords being stolen either by key loggers or shoulder-surfing attacks.
Quite surprisingly, the current bottleneck of computer security when logging for authentication is the  User Interface (UI): 
How can we enter safely secret passwords when concealed spy cameras or key loggers may be recording the login session?
Logging safely requires to design a secure Human Computer Interface (HCI) robust to those attacks.
We describe a novel method and system based on entering secret ID passwords by means of associative secret UI passwords that provides zero-knowledge
to observers. 
We demonstrate the principles using a color Personal Identification Numbers (PINs) login system and describes its various extensions.
\end{abstract}
 
\parindent 0cm
{\bf Keywords}:
HCI, password, PIN, shoulder-surfing  attack, key logger,  one time password.

\section{Introduction}

{Nowadays}, we are logging more and more to cloud services using Internet terminal in public spaces with the threat of passwords being stolen.
Stealing passwords can be done either by machine key and mouse loggers or by  human shoulder-surfing  attacks. 
Purloining passwords often opens the door to identity theft.
The threat is even more important as concealed spy or surveillance digital cameras  may be recording the login session, and later computer vision techniques may be used to recover the secret password. 
Wired magazine had in December 2012 its cover page entitled \verb|kill the P@55W0rD|
 to emphasize on the emergency of rethinking the password protection systems.
There is a strong need  not only to rethink passwords~\cite{RethinkingPasswords-2013} but also to reconsider the User Interface (UI) to input safely passwords.
This is an important Human Computer Interface (HCI) problem that is gaining more and more attention with dedicated conferences like
the  Symposium On Usable Privacy and Security (SOUPS). 

Ideally, we should not need to log in as the machine should recognize the unique ``Me'' as sophisticated computer HAL 9000 did in Kubrick's film {\it 2001: A Space Odyssey} (1968). That is, ultimately, it is not the user who shall authenticate oneself but the machine who shall authenticate the user, securely.
Although artificial intelligence has made significant progress since that 1960's movie release, we are yet far from such a robust non-invasive biometric system.\footnote{Current biometric systems include 3D head scan for example.} 
Brain machine interfaces  for logging have also recently been considered~\cite{Passthoughts-2013} with the hope, one day, of ditching pass-words and replace them with ``pass-thoughts.'' (Section~\ref{sec:concl} shall review prior work.)
 
Thus we ask ourselves what would be a {\em secure UI system} to log in public spaces in front of other people using possibly a machine with unprotected and readable input/output interfaces (I/O)? 
One solution is to use a One Time Password\footnote{\url{http://en.wikipedia.org/wiki/One-time_password}} (OTP) {\em physical token device} that is assumed to belong privately to the user. Each time the user logs in, it asks the token for a newly on-demand generated password ({\it e.g.}, typically by pushing a button on the physical token).
However, this device may be inadvertently stolen or, even worse, some OTPs  maybe generated by an attacker without stealing the device.
The latter case is far more difficult to detect for the user that has assumed to always securely carry the device!

Notice that passwords should never be stored plainly on servers. This has dramatic consequences when servers are cracked, but unfortunately happens quite often as reported in the news.
UNIX login program stores {\em encrypted} passwords either plainly in {\tt /etc/passwd} file or in a privileged-access {\tt /etc/shadow} file.
It uses the user password as the key to cypher iteratively 25 times the 64-bit 0 string using the {\tt crypt()} DES.
To strengthen even more the password a $2$-character ``salt'' is generally added~\cite{Salt-2000}.
 
Our proposed system, called {\em Color PINs}, is based on using an associate UI password for entering safely a secret ID password.
The system only requires to have a secure login software ({\it e.g.}, signed with a cryptographic key for avoiding Trojan attacks) and a random generator.
 
The paper is organized as follows: 
Section~\ref{sec:colorPIN} describes the   UI login system and explain its fundamental principles.
It is followed by Section~\ref{sec:ext} that presents some of its extensions.
Finally, Section~\ref{sec:concl} reviews prior work and discusses on several perspectives.

\section{The color PIN system\label{sec:colorPIN}}

We propose a simple technique to login using a public login name based on {\em two associated passwords} instead of the traditional single secret password scheme.
Since it is easy to record insidiously keyboard key strokes, our logging system uses a  graphics keyboard and a  cursor board  User Interface (UI).

\subsection{Description of the basic system}

\begin{figure}
\centering
 \begin{tabular}{cc}
 \includegraphics[bb=0 0 1270 677, width=0.48\textwidth]{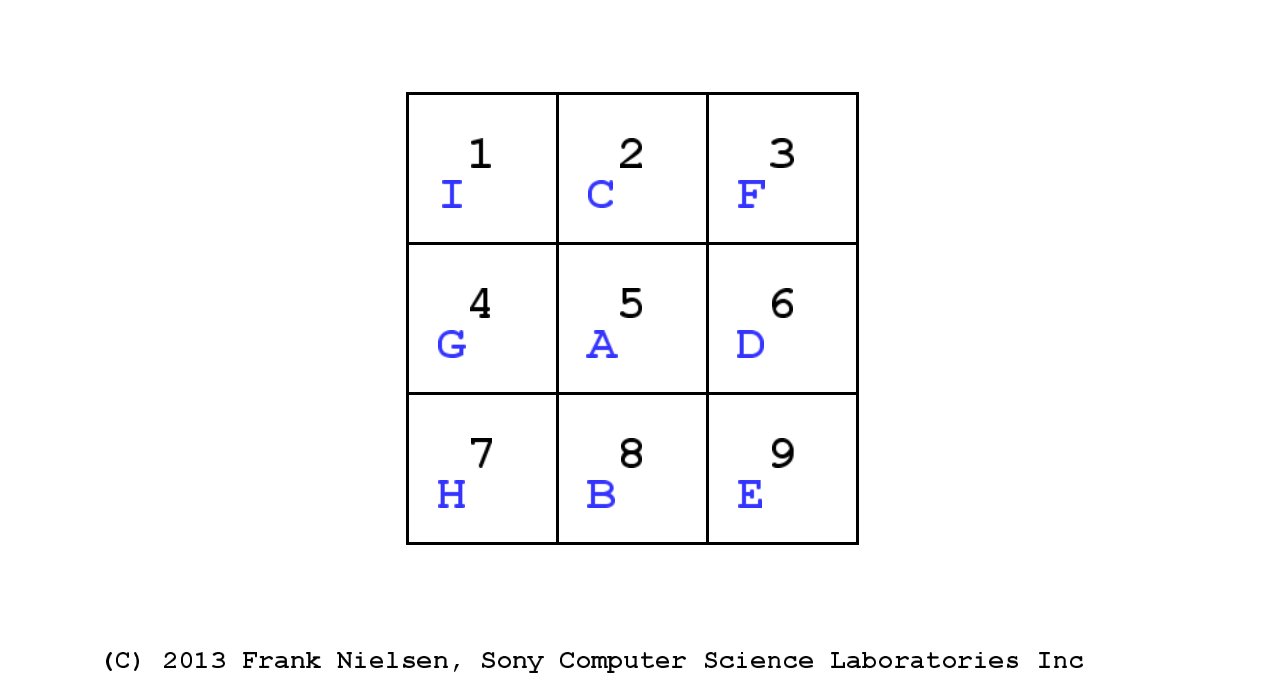}  &
 \includegraphics[bb=0 0 1272 673, width=0.48\textwidth]{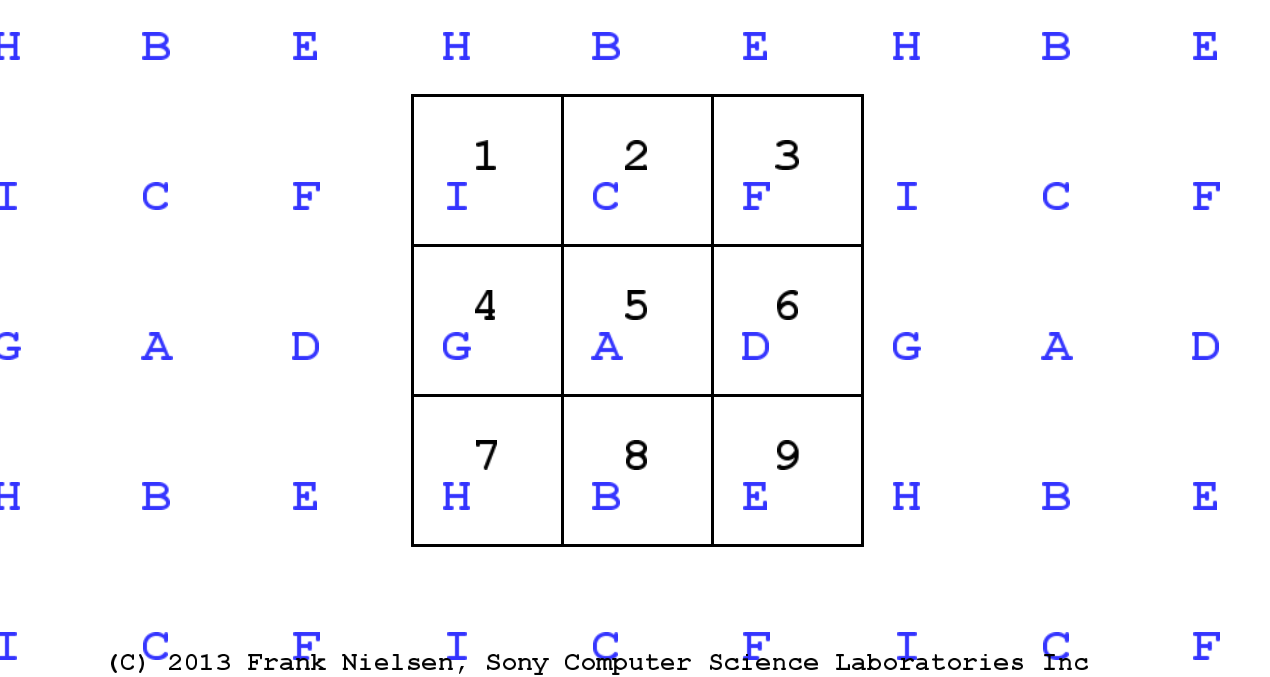}\\
 (a) & (b)
 \end{tabular}
  
\caption{
Snapshots of the system: The moveable  $3\times 3$ cursor letter board   is displayed on top of the fixed digit keyboard.  
To simulate the 2D torus topology of the cursor keyboard, we display $3\times 3$ translated (Euclidean) copies (b) of the letter cursor board with offset the board dimension, and clip the rendering to the fixed digit keyboard (a). 
\label{fig:torus}}
\end{figure}

Once a user enters his/her login name, the system receives a {\em UI password}.
The UI password can be defined and kept secret by the user beforehand, or generated on-line using the user profile as described later in Section~\ref{sec:userprofile}.
On the graphical screen, we display a fixed {\em key board} and allows interaction using a second  moveable {\em cursor board} of identical dimension.
This is a {\em two-layered} display.
For example, the fixed keyboard can be a $3\times 3$ array\footnote{For sake of simplicity, we choose 2D arrays in this presentation as we are usually accustomed with this numeric pad layout. However, the array can also be  1D (linear), or even 3D (volumetric) if one wishes.} of digits labeled from $1$ to $9$, and the cursor board can be a $3\times 3$ array of letters labeled from $A$ to $H$. 
Let $n=9$ denote the number of keys.
The moveable cursor board is controlled with a mouse or touch panel finger interaction.
To enter his/her secret password, the user needs to align (a task performed in his/her brain) the cursor key with the {\em corresponding} digit key and confirm that input using a mouse click or double finger tap event, for example. 
The mouse cursor of the operating system is preferably hidden, but this is not a strict requirement.
The moveable cursor array, which is controlled by interaction, overlays on the fixed digit keyboard and warps on the fixed board edges to implement the 2D torus topology.
This is depicted in Figure~\ref{fig:torus}.
Furthermore, to scroll endlessly with the 2D torus topology, we implement screen edge warping of the OS mouse cursor.
Notice that in practice, it is enough to display $3\times 3$ translated copies of the moveable cursor board with offsets the dimension of the board, as depicted in Figure~\ref{fig:torus}.
There is no way for an observing intruder to detect {\em which} key is aligned with which cursor as all pairs are visually matching.
Only the user and the logging system know which key digit-letter pair shall be selected when input validation is triggered.
Thus an observer needs to memorize the $n$ potential pairs at each step if he/she wants to break the password.
For passwords of length $k$, this ends up with an exponential number, $n^k$ combinations, of potential passwords to test to crack the user secret password.
Each time the user enters a key, an asterisk is displayed to notify the event and both the fixed key  and the moveable cursor  boards are randomly shuffled, thus yielding no correlation for the next digit-letter alignment task.
Users can reset or validate the last entered key using either mouse middle or right clicks, or by touching the screen at virtual reset/validate button locations.

\begin{figure}
\centering

\begin{tabular}{cc}
\includegraphics[bb=0 0 458 476, width=0.5\textwidth]{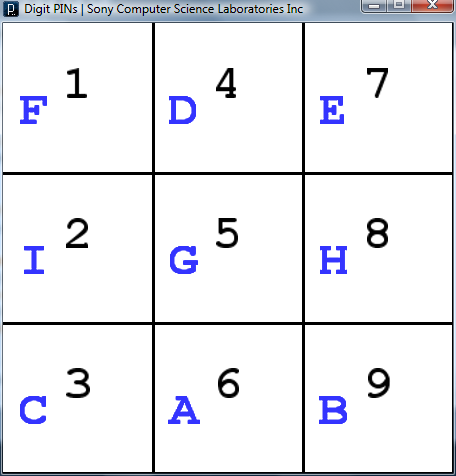} & \includegraphics[bb=0 0 458 476, width=0.5\textwidth]{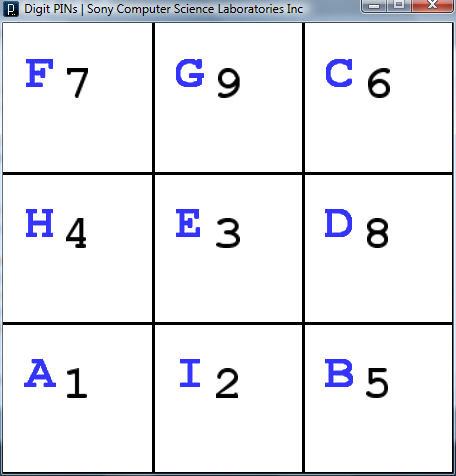} \\
\underline{3}141 & 3\underline{1}41 \\
\underline{C}AHB & C\underline{A}HB \\ 
\includegraphics[bb=0 0 458 476, width=0.5\textwidth]{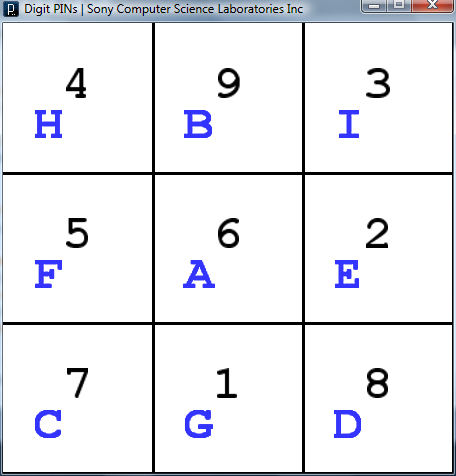} & \includegraphics[bb=0 0 458 476, width=0.5\textwidth]{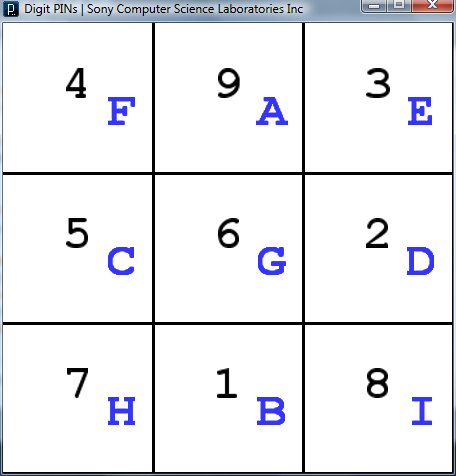} \\
31\underline{4}1 & 314\underline{1} \\
CA\underline{H}B & CAH\underline{B} \\ 
\end{tabular}
 
\caption{\label{fig:session}Login session for the associative password system using a $3\times 3$ fixed digit board with a   $3\times 3$ moveable letter cursor. The user enters sequentially the password $3141$ using the UI password  $CAHB$. After each digit input is confirmed by a mouse click, both boards are randomly shuffled.}
\end{figure}

To further be resistant to mouse logger that could record and replay later the mouse events, we choose the origin of the moveable cursor pointer randomly inside the fixed board. This origin is used when displaying the UI controlled moveable cursor. 

Notice that instead of displaying the full moveable cursor board with $n$ keys on top of the fixed graphics board, it is enough to display at least $l\geq 2$ of them including the correct associative key in order to get an exponential number, $l^k$, of potential combinations. 

Figure~\ref{fig:session} depicts the screen captures of a session for a fixed digit-moveable $3\times 3$ letter board implementation of the system:
 The user enters the ID digit password {\tt 3141} using the associated UI letter password {\tt CAHB}. 
 Note that after each digit is entered by mouse clicking both the digit and letter boards are shuffled randomly.

\begin{figure}
\centering
\includegraphics[bb=0 0 1000 452, width=0.5\textwidth]{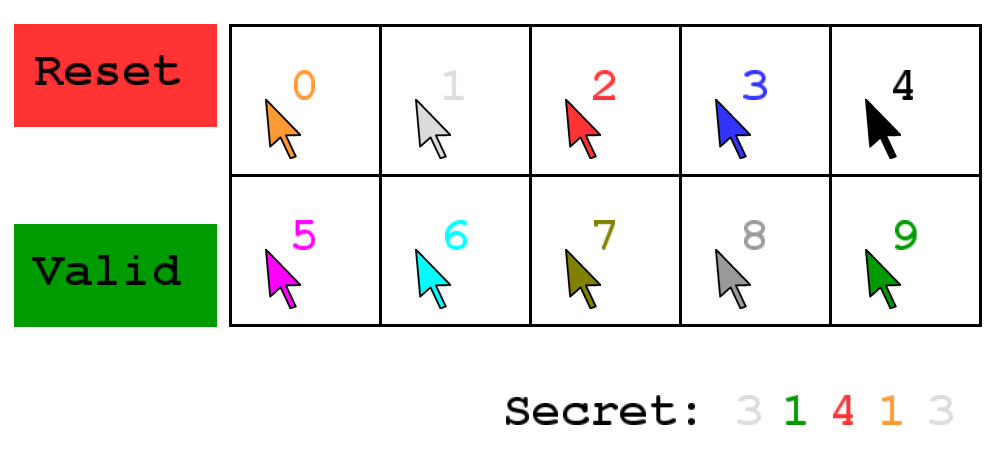}

\caption{\label{fig:colorPIN}The Color PIN system using a $2\times 5$ color layout. The digits are colored with the corresponding matching cursor color.}
\end{figure}

Figure~\ref{fig:colorPIN} shows a similar system using a fixed board with a moveable color cursor board. The layout is $2\times 5$ with the following
 moveable color cursor board:\\
 \begin{center}
 \begin{tabular}{ccccc}
 BLACK & ORANGE & LIGHTGRAY& RED& BLUE\\
  GREEN& PURPLE & AQUA & OLIVE & GRAY
  \end{tabular}
  \end{center}
The digits are colored with their respective cursor color currently covering the digit key.
The secret password is 31413 and that password should be entered with the following color cursor combination:\\

\begin{center}
 \begin{tabular}{ll}
 Color cursor (UI PWD) & Digit cursor (ID PWD) \\ \hline\hline
 LIGHTGRAY& 3\\
  GREEN& 1 \\
   RED& 4\\
    ORANGE& 1\\
     LIGHTGRAY & 3
     \end{tabular}
     \end{center}

Using the same secret ID and UI passwords at several login sessions weakens the security.
Indeed, if no shuffling of the digit and/or cursor board is done, then recording, say, at the first position of the '1' digit the color sequence will provide the password information to observers. When shuffling (at least one of the two boards), we make different permutations of digits-color cursors.
However, video recording several login sessions and analyzing the common matching digit-color pairs will provide information on the passwords.
Thus we should ideally change the UI password after at each session. This can be done by using a OTP token device for generating UI password keys.
The user then keeps his/her secret password and generate a UI password when logging. The advantage of the color PIN system over the traditional OTP token device is that even if the physical token is stolen then it does not allow to properly log in as the ID password is not compromised.

\subsection{Implementation details}
The color PIN system has been implemented using the {\tt Processing} language (\url{processing.org}), version 2.0b8.
Wrapping the cursor operating system on the screen edges (for endless smooth toric motion) is done using by using Java\texttrademark{} {\tt Robot} class and {\tt mouseMove} function.

\section{Some extensions of the color PINs system\label{sec:ext}}

We present several extensions of the associative PIN code system.

\subsection{Using legacy password systems}

We can use legacy systems based on a single PIN by decomposing it into two parts.
For sake of simplicity, consider the pin size of even length $2k$.
Then the first $k$ digits of the PIN are used for the traditional ID password and the last $k$ digits are used for the UI password with the mapping $1\rightarrow A$, ..., $9\rightarrow H$ when considering digit-letter boards, or $1\rightarrow$ BLACK, ..., $9\rightarrow$ GRAY when considering digit-color boards.
Thus we can enter a password of length $2k$ on legacy system using only $k$ association pairs.
For example, the PIN $1234$ is entered using $C$ and $D$ cursors aligned to keys $1$ and $2$, respectively (or using LIGHTGRAY and RED color cursors).
Note that by doing so, we weaken the security as we have $n^{k}$ potential combinations instead of $n^{2k}$ in a login session.
In general, we can create a secure password including the UI password by juxtaposing the two parts: \fbox{ID PIN}+\fbox{UI PIN}.
The UI PIN may be shorter than the PIN. For example, we can use letter 'A' cursor to enter the full PIN. 
When the UI PIN is shorter than the ID PIN (e.g., $1234AB$), we choose the current cursor cursor using the {\em modulo} operator (here, '1' and '3' should be entered using 'A' and '3' and '4' using 'B').

\subsection{Graphics board skins\label{sec:skin}}

\begin{figure}
\centering
 \includegraphics[bb=0 0 453 453, width=0.45\textwidth]{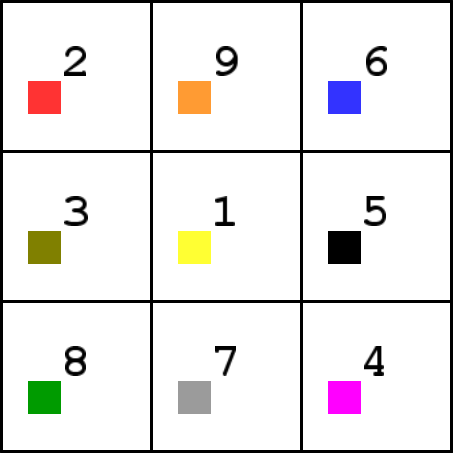}  
 \includegraphics[bb=0 0 453 453, width=0.45\textwidth]{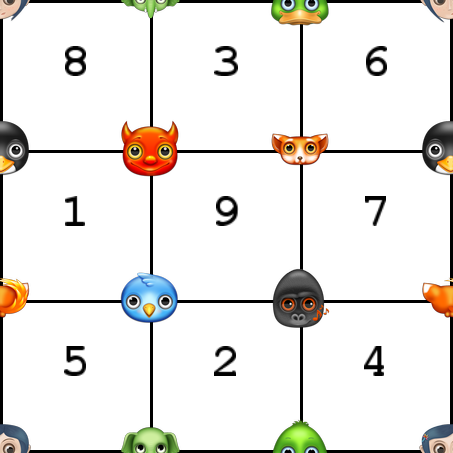} 
\caption{
Snapshots of the login system using different rendering styles for the moveable cursor board: (left) color boxes and (right) icons. 
\label{fig:skins}}
\end{figure}

To make the system more general, we just need to define two boards of $n$ elements: the fixed board elements $F_1, ..., F_n$ and the moveable cursor board elements $M_1, ..., M_n$.
Those elements are numbered with function $N(\cdot)$ universally using numbers $1$, ..., $n$: $N(F_i) = N(M_i)$.
This allows for various rendering styles.
For example, we can use vacation images\cite{PassImage-2012} for the fixed board and numbers for the cursor board, etc.
We can use icon sets or color box sets as two different rendering styles for the moveable cursor array as depicted in Figure~\ref{fig:skins}.

Instead of using 2D graphics boards, 1D layouts of fixed and moveable boards may be preferred for, say, personal picture browsing, and require less graphics space to display for a login session.
Thus this kind of ribbon user interface maybe appropriate to display at the bottom of other software when logging is required.
We may also add {\em visual effects}: For example, we may have a color PIN to memorize, and have the color cursor rolling overs the digits.
Each time the color is on the digit key, all digits are colored with their corresponding colors (as shown in Figure~\ref{fig:colorPIN}). 
Indeed, Human tend to memorize visual objects rather easily compared to juxtaposed ID-UI associated PINs.
The moveable cursor may be textured (like bark, grass, or street road textures) and the digits textured with the corresponding highlighted texture cursor.

\subsection{UI password generation based on user profile\label{sec:userprofile}}

To help user memorize the UI password and generate many UI passwords, we define a {\em user profile} by asking a set of $k$ questions, each with $n$ choices like what is her favorite food, favorite place, favorite color, favorite celebrity, favorite movie, favorite music, etc.
Each time the user enters a key, the moveable cursor skin changes to the next mode (e.g., food$\rightarrow$place$\rightarrow$color$\rightarrow$celebrity$\rightarrow$movie$\rightarrow$music, etc.).
Furthermore, for $k$-length passwords, we generate a random permutation on the question orders (yielding $k!$ UI passwords) so that the user knows which moveable cursor shall be used to enter the current ID PIN digit.
This profiling frees users from memorizing   UI passwords but may be less secure when observers know or guess his/her preferences.

\subsection{Cursor control using another device}

The moveable cursor board may be controlled using events sent from another device.
For example, a tablet or a game pad with a touch screen may be used to send event relative displacement orders to the login application.
The orders can be absolute coordinates (rescaled to fit the display screen) or relative motion orders encoding a translation.
Note that in practice, it is interesting to have a board display of relatively small size while providing a large UI pad for precise interaction.
This makes the input more precise for users, say, on  smart phones equipped with touch screens, where users can touch anywhere on the screen for controlling the moveable cursor array on the fixed board.

\section{Prior work and discussion\label{sec:concl}}
We first present prior work related to our color PIN system (see references therein too):
The CursorCamouflage~\cite{CursorCamouflage-2012} system   displays a set of {\em dummy} cursors that makes it difficult for observers to correlate with the user hand motion. However, a careful video analysis of the recorded login session by advanced computer vision techniques may decipher the secret key by analyzing hand-mouse cursor correlations. 
In comparison, our color PIN system as no dummy cursor but instead use two key boards and two PINs to identify oneself. In~\cite{UIConvexHullClick-2006}, the authors present a shoulder-surfing resistant UI for entering password that uses {\em pass-icons} blended with other icons on a 2D layout; The user is required to pass {\em several challenges} where each challenge asks to click inside the convex hull of the pass icons. This system (like ours) does not require to click on the pass icons. It is however more time consuming and requires laymen to be familiar with the convex hull definition.
In~\cite{PINEntry-2004}, the authors describe {\em cognitive trapdoor games }where the users has two select on which set the current PIN code digit is contained.
After a few selections, the system knows by ``intersecting'' the challenge subsets which digit was entered, and proceed for entering the next digit, etc. One major drawback is that it provides limited resilience when the session login is video-taped.
The closest work to our color PIN system may be found in~\cite{Fakecursor-2008}.\footnote{\url{http://www.netaro.info/~zetaka/projects/fakePointer/index.html}} 
The FakeCursor system manages a fixed secret and a disposal secret:
He/She enters her pin code by aligning the  secret digit on the fixed disposal icons using left/right ATM-like arrow buttons.  
We can interpret FakeCursor as a discrete UI working on the 1D ring topology.
In comparison, our system is fully graphical as it uses 2D torus topology and requires much less user input.  
Our system has also higher-level abstraction as it requires to define two boards generically that are associated using a universal numbering (see Section~\ref{sec:skin} on skins), or uses a color UI OTP token device for entering the ID password.

The main drawback of the color PIN system is to require memorizing additional information: Namely, the UI password. 
Another potential threat is by using gaze tracking and advanced computer vision, it might be possible to guess which part were ``intentionally''  aligned by observing the user' eyes (a kind of computational mentalism). This risk is minimized by showing a small board size.
Interestingly, note that conversely, there are login systems that have been designed based on eye gaze input~\cite{GazedBasedPWD-2007}. 

We have presented a login system that considers {\em zero-knowledge UI} for entering passwords.
In perspective, instead of using the (parallel) visual perception system for entering a ID password, we could use  tactons~\cite{PINTacton-2010}: 
The shuffled digits on the digit boards shall not be visually displayed, but tactilically rendered using tactons, and the user shall press the corresponding tacton corresponding to the secret PIN. This system is particularly well-suited for visually impaired people.

\end{document}